\documentclass[a4paper,12pt]{article}
\pdfoutput=1
\usepackage{graphicx}
\usepackage{subcaption}
\usepackage[margin=1in]{geometry}
\usepackage{afterpage}
\usepackage{youngtab}
\usepackage{amsmath}
\usepackage{color}
\usepackage{simplewick}

\newcommand{\be}{\begin{equation}}
\newcommand{\bea}{\begin{eqnarray}}
\newcommand{\ee}{\end{equation}}
\newcommand{\eea}{\end{eqnarray}}
\newcommand{\Tr}{{\rm Tr}}
\def\Xint#1{\mathchoice
   {\XXint\displaystyle\textstyle{#1}}%
   {\XXint\textstyle\scriptstyle{#1}}%
   {\XXint\scriptstyle\scriptscriptstyle{#1}}%
   {\XXint\scriptscriptstyle\scriptscriptstyle{#1}}%
   \!\int}
\def\XXint#1#2#3{{\setbox0=\hbox{$#1{#2#3}{\int}$}
     \vcenter{\hbox{$#2#3$}}\kern-.5\wd0}}

\def\dashint{\Xint-}

\begin{document}
\makeatletter
\@addtoreset{equation}{section}
\makeatother
\renewcommand{\theequation}{\thesection.\arabic{equation}}
\vspace{1.8truecm}

{\LARGE{ \centerline{\bf Holography for Tensor models}}}  

\vskip.5cm 

\thispagestyle{empty} 
\centerline{ {\large\bf 
Robert de Mello Koch$^{a,b,}$\footnote{{\tt robert@neo.phys.wits.ac.za}}, 
David Gossman$^{b,}$\footnote{{\tt  dmgossman@gmail.com}},
Nirina Hasina Tahiridimbisoa$^{b,c,}$\footnote{{\tt nirina@aims.ac.za}} }}\par
\vspace{.2cm}
\centerline{{\large\bf and
Augustine Larweh Mahu${}^{b,d,}$\footnote{ {\tt aglarweh@gmail.com}} }}

\vspace{.4cm}
\centerline{{\it ${}^{a}$ School of Physics and Telecommunication Engineering},}
\centerline{{ \it South China Normal University, Guangzhou 510006, China}}

\vspace{.4cm}
\centerline{{\it ${}^{b}$ National Institute for Theoretical Physics,}}
\centerline{{\it School of Physics and Mandelstam Institute for Theoretical Physics,}}
\centerline{{\it University of the Witwatersrand, Wits, 2050, South Africa}}

\vspace{.4cm}
\centerline{{\it ${}^{c}$ Institute of High-Energy Physics of Madagascar (iHEPMAD)}}
\centerline{{\it University of Ankatso, Antananarivo 101, Madagascar,}}

\vspace{.4cm}
\centerline{{\it ${}^{d}$ Department of Mathematics,,}}
\centerline{{\it  University of Ghana, P. O. Box LG 62, Legon, Accra, Ghana.}}

\vspace{1truecm}

\thispagestyle{empty}

\centerline{\bf ABSTRACT}

\vskip.2cm 
In this article we explore the holographic duals of tensor models using collective field theory. 
We develop a description of the gauge invariant variables of the tensor model.
This is then used to develop a collective field theory description of the dynamics.
We consider matrix like subsectors that develop an extra holographic dimension.
In particular, we develop the collective field theory for the matrix like sector of an interacting tensor model.
We check the correctness of the large $N$ collective field by showing that it reproduces the perturbative
expansion of large $N$ expectation values. 
In contrast to this, we argue that melonic large $N$ limits do not develop an extra dimension.
This conclusion follows from the large $N$ value for the melonic collective field, which has delta function support.
The finite $N$ physics of the model is also developed and non-perturbative effects in the $1/N$ expansion are exhibited.

\setcounter{page}{0}
\setcounter{tocdepth}{2}
\newpage
\setcounter{footnote}{0}
\linespread{1.1}
\parskip 4pt

{}~
{}~

\section{Introduction}

In the last 20 years genuinely new and fascinating insights into the large $N$ expansion have been achieved.
The ribbon graph expansion proposed by 't Hooft suggested a deep connection between non-Abelian gauge theories and 
string theories more than 40 years ago.
The expansion identifies the surface triangulated by the ribbon graph with a string worldsheet\cite{tHooft:1973alw}.
This initial proposal has found beautiful confirmation in the duality between the large $N$ expansion of ${\cal N}=4$ 
super Yang-Mills theory and the loop expansion of IIB string theory on asymptotically AdS$_5\times$S$^5$ 
spacetimes\cite{Maldacena:1997re,Gubser:1998bc,Witten:1998qj}.  
This connection goes by the name of holography, or the gauge theory/gravity duality and for any theory with adjoint 
valued variables, one expects a duality with a string theory.
The large $N$ limits of vector models are much simpler.
There are fewer gauge invariants that one can form and the large $N$ limit of these models is explicitly solvable.
Remarkably, they are also equivalent to higher dimensional theories of gravity\cite{Sezgin:2002rt,Klebanov:2002ja}, 
but in this case the relevant gravity theories are the so-called higher spin theories\cite{Vasiliev:2003ev,Vasiliev:1999ba}.
In the light of this progress, it is natural to work out the large $N$ expansion and holography for tensor models.
This is the main motivation for our study.

Tensor models have recently gained popularity, as close cousins of the SYK model\cite{Sachdev:1992fk,Kitaev}.
The SYK model is a quantum mechanical model of Majoranna fermions, interacting with a random coupling.
The model is fascinating because it develops an emergent conformal symmetry in the IR where it is strongly 
coupled\cite{Maldacena:2016hyu}.
Its strongly coupled large $N$ limit has been constructed exactly and used to demonstrate that the model saturates the chaos 
bound\cite{Maldacena:2015waa}.
These features strongly suggest that the model is holographically dual to a black hole.
The tensor models share many of these features with one significant advantage: they have no 
disorder\cite{Witten:2016iux}. 

Tensor models are rather interesting in their own right.
Tensor models are asymptotically free\cite{BenGeloun:2012pu,BenGeloun:2012yk,Rivasseau:2015ova}.
The large $N$ limit of tensor models involves specific discretizations of the $d$-sphere, known as melonic 
\cite{Bonzom:2011zz}, which provide an analytic description of the universality class of the branched polymer phase of 
Euclidean dynamical triangulations\cite{Ambjorn:1990wp}. 
Further, the multi-critical behaviors \cite{Bonzom:2012hw} can be interpreted as critical, non-unitary matter 
\cite{Bonzom:2012qx,Bonzom:2012np}, just like in 1-matrix models \cite{Kazakov:1989bc}.
Our main focus is on the holography of tensor models.
For related studies, see \cite{deMelloKoch:2017bvv,Vasiliev:2018zer,Delporte:2018iyf}, as well as
\cite{Carrozza:2015adg,Klebanov:2016xxf,Klebanov:2017nlk,Bulycheva:2017ilt,Giombi:2018qgp,Klebanov:2018fzb,Ferrari:2017jgw,Pascalie:2018nrs,Benedetti:2019eyl}.

One possible approach to the holography of tensor models, is to express the dynamics of the model in terms of gauge 
invariant variables.
A systematic method to approach this problem is the collective field theory\cite{Jevicki:1979mb,Jevicki:1980zg}.
The collective field theory for vector models expresses the dynamics in terms of a bilocal field, which has a very direct 
connection to the dual higher spin gravity in AdS 
space\cite{Das:2003vw,Koch:2010cy,Jevicki:2011ss,Das:2012dt,Koch:2014aqa}.
The bilocal description gives a higher dimensional theory, with $1/N$ as the loop expansion parameter.
A similar analysis for single matrix dynamics can be carried out using the eigenvalue density as the gauge invariant variable.
This gives the Das-Jevicki Hamiltonian, which reproduces the dynamics of the $c=1$ string\cite{Das:1990kaa}. 
The resulting collective field theory is higher dimensional and it too has $1/N$ as the loop expansion parameter. 
For multi-matrix models it is hard to parameterize the loop space of gauge invariants in a useful way and this has proved to be
a central obstacle to deriving collective field theory.
Since there are suggestions that tensor models are richer than vector models, but perhaps simpler than matrix models, it is
worth examining their collective description.
This is our main goal in this article.
If an example can be found, with a space of invariant variables that is richer than vector models, it may shed important
insights into how to tackle the space of invariant variables for multi-matrix models.

Since collective field theory plays a central role in our study, we give a brief review of the method in Section \ref{Collective}.
This review includes a discussion of how the Jacobian for the change to invariant variables is computed and then how it
determines the collective Hamiltonian.
Section \ref{Invariants} discusses one approach towards enumerating and then constructing the gauge invariants of
tensor models. 
We focus on rank 3 tensors which transform in the fundamental of $U(N_1)\times U(N_2)\times U(N_3)$.
This space of invariants is enormous and hard to handle.
We start by constructing invariants under $U(N_1)\times U(N_2)$, which can be managed.
The result is a collection of $U(N_3)$ tensors of arbitrary even rank, which is
a definite intermediate result on the path towards characterizing the complete space of invariants.
Section \ref{Strings} looks for and identifies sectors that are dynamically decoupled.
These correspond to stringy states and indicate that there are strings in the Hilbert space of the tensor model and 
further, that this is a useful language for classes of questions that can be pursued.
We explicitly demonstrate how collective field theory reproduces correlators in a model with interactions.
This physics recovers many features expected of matrix models.
In an attempt to find some truly tensor model physics, Section \ref{Melons} considers the construction of the collective
field theory of melonic variables.
Using certain known results for the melonic variables, we are able to construct the leading large $N$ descriptions of
the melonic invariant variables. We find that there is no holographic dimension emergent at large $N$.
In Section \ref{Expansions} we use a complete basis of the tensor model invariants provided by the restricted
Schur polynomials, to investigate non-perturbative contributions to the large $N$ expansion.
We end with conclusions in Section \ref{Conclusions}.

\section{Collective Field Theory}\label{Collective}

The basic method we employ in this paper is the collective field theory\cite{Jevicki:1979mb,Jevicki:1980zg}.
Our goal in this section is to give a general review of the method. 
Collective field theory provides a systematic construction of the dynamics of invariant observables of the theory. 
This is typically accompanied by the emergence of extra holographic dimensions as well as a reorganization of the dynamics
such that $1/N$ is the new loop counting parameter.
It is therefore the ideal framework with which to pursue the holography of tensor models and to provide insight into
the structure of the large $N$ expansion for these models.
Collective field theory achieves a direct change of variables to the invariant observables, taking careful account of the Jacobian
associated to this change of variables. 
The Jacobian is a highly non-trivial functional of the fields and it produces a non-linear (collective) Hamiltonian which
provides a complete description of the theory.
The actual set of invariant variables relevant for the tensor model will be described in Section \ref{Invariants}. 

We will begin with some comments to orient the reader who may not be familiar with the details of collective field theory.
The method begins by identifying a suitable set of invariant observables.
Rewriting the Hamiltonian of the theory in terms of the invariant variables entails using the chain rule to 
express the kinetic energy in the new variables.
After the rewriting the Hamiltonian is no longer explicitly hermittian.
This is of course a consequence of the fact that the change of variables is accompanied with a non-trivial Jacobian $J$ and
consequently, there is a non-trivial measure on the space. 
By rescaling all states and operators as follows
\bea
|\Psi\rangle\to J^{1\over 2}|\Psi\rangle\qquad
\hat{O}\to J^{1\over 2}\hat{O}J^{-{1\over 2}}
\eea
one trivializes the measure.
Thus after the transformation the Hamiltonian has to be manifestly hermittian. The equation $H-H^\dagger=0$ then
implies a differential equation that determines the Jacobian.
Before the rewriting, $N$ counts the number of dynamical variables of the system.
For example, we may consider a vector model where the field $\phi^a$ has $N$ components $a=1,2,\cdots,N$ or
a matrix that has $N$ eigenvalues.
After the rewriting, $N$ simply appears as a parameter of the theory and this greatly facilitates carrying out the large $N$
expansion.

Consider a system with Hamiltonian 
\bea
H = -{1\over 2}\sum_{i=1}^N {\partial\over\partial x_i}{\partial\over\partial x_i}+ V (x_i) 
\eea
Here $x_i$ are the original variables of the system.
Denote the invariant field variables by $\phi_{C}$ where $C$ is an index for the invariants.
In the large $N$ limit, the change of variables from the original variables to invariants is a reduction in the number
of degrees of freedom. 
These variables are treated as independent in the large $N$ limit. 
This last point is highly non-trivial: one might have expected that the transformation only makes sense if the
number of old and new variables are the same.
This is clearly not the case.
We might for example be changing from $N$ eigenvalues to a density of eigenvalues $\phi (x)$ that now has a continuous
parameter $x$.
This simply implies that not all $\phi (x)$ are independent.
There are further constraints that arise at finite $N$ related to the stringy exclusion 
principle\cite{Jevicki:1998rr,McGreevy:2000cw}.
Convincing evidence coming from explicit computations in a number of 
examples \cite{Demeterfi:1991tz,Demeterfi:1991nw,Jevicki:1980zq,Ishibashi:1993nqz} suggest
that we should treat the collective variables as independent when formulating the perturbative large $N$ expansion.
Our computations in this article add to this list of evidence.

Consider changing variables from the original variables to a set of gauge invariant variables. 
The kinetic term becomes
\bea
T = - {1\over 2}\sum_{i=1}^N {\partial\over\partial x_i}{\partial\over\partial x_i} = 
-{1\over 2}\sum_{C,C'} \Omega (C,C') {\partial\over\partial \phi_C}{\partial\over\partial {\phi}_{C'}} +
{1\over 2}\sum_{C}  \omega (C) {\partial\over\partial {\phi}_{C}} 
\eea
where
\bea
\Omega (C,C') = \sum_{i=1}^N {\partial\phi_C\over\partial x_i}{\partial{\phi}_{C'}\over\partial x_i}
\qquad\qquad
\omega (C) = - \sum_{i=1}^N {\partial^2\phi_{C}\over\partial x_i\partial x_i}
\eea
$\Omega (C,C')$ ``joins" invariant variables so that we write schematically 
\bea
   \Omega (C,C') =  \sum \phi_{C+C'}
\eea
where the sum runs over all possible ways of ``joining'' $\phi_C$ and $\phi_{C'}$ to produce $\phi_{C+C'}$.
Similarly, $\omega$ ``splits" loops so that 
\bea
{\omega (C) = \sum  \phi_{C'}\phi_{C''} }
\eea
where the sum runs over all possible ways of splitting the word $C$ into $C'$ and $C''$.
This change of variables is accompanied by a nontrivial Jacobian. 
This Jacobian will be the source of the non-linear interactions of collective field theory.
It ensures the unitarity of the new description. 
Indeed, after the similarity transformation 
\bea
      {\partial\over\partial {\phi}_{C}} \to 
             J^{1/2} {\partial\over\partial {\phi}_{C}}  J^{-{1/2}} =
 {\partial\over\partial {\phi}_{C}} - {1\over 2}{\over}
{\partial\ln J\over\partial {\phi}_{C}} 
\eea
one requires that the Hamiltonian is explicitly hermittian, which implies a differential equation for the Jacobian, which
determines the Hamiltonian in terms of the invariant variables. 
The differential equation is
\bea
 -\sum_{C'} (\partial_{C'} \ln J) \Omega (C',C) = \omega (C) +\sum_{C'}\partial_{C'}\Omega (C',C)
\eea
The collective field Hamiltonian is now obtained as follows
\bea
H=-\sum_{C,C'}\Omega(C,C') \left[{\partial\over \partial\phi_C}-{1\over 2}{\partial\ln J\over\partial \phi_{C}}\right] 
\left[{\partial\over \partial\phi_{C'}}-{1\over 2}{\partial\ln J\over\partial {\phi}_{C'}}\right]+V        
\eea
Using the differential equation for the Jacobian, this can be simplified as follows
\bea
H=\sum_{C,C'} \Pi(C) \Omega(C,C') \Pi(C') +
{1\over 4} \sum_{C,C'} \omega(C) \Omega^{-1}(C,C')\omega(C') + V + \Delta H
\eea
where 
\bea
\Pi(C) = - i {\partial\over \partial\phi(C)}
\eea
and
\bea
\Delta H =-\sum_C{1\over 4}\,{\partial\omega(C)\over\partial \phi_C}+{1\over 4}\,\sum_{C,C',C''}
{\partial\Omega(C^{''},C')\over\partial \phi_{C^{''}}}\,\Omega^{-1}(C',C)\omega (C)\cr\cr
+{1\over 4}\,\sum_{C,C',C'',C'''}{\partial\Omega(C^{''},C)\over \partial \phi_{C^{''}}}\,\Omega^{-1}(C,C')\,
{\partial\Omega(C',C^{'''})\over\partial\phi_{C^{'''}}}
\eea

The key obstacle in applying collective field theory, lies in handling the space of invariant variables.
The set of invariants for both multi matrix models and tensor models is large and, consequently, difficult to handle.
If one starts with a truncated set of invariants $C$ and their conjugates $C'$, through the process of ``joining" 
(contained in $\Omega (C,C')$) one typically generates new loops not in the original set. 
This makes it hard to find simpler sub sectors whose dynamics can be studied. 
For relevant related work on this issue, that we use in this study, 
see \cite{deMelloKoch:2002nq,deMelloKoch:2003pv,Donos:2005vm}.

Applying the collective field theory to matrix quantum mechanics leads to a highly non-trivial
string field theory that beautifully captures the dynamics of the $c=1$ string\cite{Das:1990kaa}.
Our primary motivation in this paper is to initiate the construction of the the analogous theory of the invariants 
of the tensor model quantum mechanics.  

\section{Invariant Variables}\label{Invariants}

In matrix models the basic gauge invariant observables are built from a matrix $Z_b^a$ which transforms as
$Z\to U^\dagger Z U$ under a gauge transformation. The gauge invariant observables are given by traces and products of traces.
 For theories with more than one matrix, traces are populated with arbitrary products of the different matrices.
Although this space of invariants is complicated, many results have been established.
We know how to count gauge invariant operators both at infinite and at finite $N$ for any number of 
matrices\cite{Sundborg:1999ue,Aharony:2003sx,Gaiotto:2012xa,Kim:2012gu,Dolan:2007rq,Koch:2012sf,Pasukonis:2013ts}.
There is a systematic $1/N$ expansion for correlation functions, 
the expansion is phrased in terms of ribbon graphs and corrections to the planar limit come in powers of ${1\over N^2}$.
The expansion has a graphical meaning as a sum over surfaces (string worldsheets)\cite{tHooft:1973alw}.
Our goal to generalize as much of this description as possible to tensor models.

In the tensor model we consider, the basic field is $\phi^{abc}$ and its conjugate $\bar\phi_{abc}$.
The gauge group is $U(N_1)\times U(N_2)\times U(N_3)$.
Under a gauge transformation the fields transform as
\bea
\phi^{abc}\to (U_1)^a_d (U_2)^b_e (U_3)^c_f\phi^{def}\qquad \qquad
\bar\phi_{abc}\to (U_1)_a^d (U_2)_b^e (U_3)_c^f\bar\phi_{def}
\eea
where $U_i\in U(N_i)$. The complete set of gauge invariant operators is enormous and making sense
of this space of invariants is the basic obstacle that we must overcome. Our approach is to deal with
the problem in two steps: first construct the gauge invariants under $U(N_1)\times U(N_2)$, and then deal with the 
last gauge symmetry as the second step. The advantage of breaking the problem up in this way is that the first step
can be tackled: there is a simple solution that can be written down explicitly. The set of $U(N_1)\times U(N_2)$ invariants
is spanned by the $U(N_3)$ tensors given by
\bea
T(2n)^{i_1i_3\cdots i_{2n-1}}_{i_2 i_4\cdots i_{2n}}=\phi^{a_1b_1i_1}\bar\phi_{a_1b_2i_2}
\phi^{a_2b_2 i_3}\bar\phi_{a_2b_3 i_4}\cdots \bar\phi_{a_n b_1 i_{2n}}\label{newobs}
\eea
To prove that these span the $U(N_1)\times U(N_2)$ invariants, we will count the number of $U(N_3)$ invariants
and then compare to an independent count.

Our first task is to provide the independent count of the number of $U(N_1)\times U(N_2)\times U(N_3)$ invariants.
This problem has been solved in \cite{Geloun:2013kta,deMelloKoch:2017bvv}.
Invariants constructed using $n$ $\phi^{abc}$s and $n$ $\bar\phi_{abc}$s are in one-to-one
correspondence with elements of the double coset
\bea
 S_n\setminus S_n\times S_n\times S_n\, /\, S_n
\eea
where $S_n$ is the symmetric group.
The number ${\cal N}_3$ of invariants, which is equal to the order of this double coset is given by
\bea
{\cal N}_3 ={1\over (n!)^2}\sum_{\sigma_1,\sigma_2,\sigma_3\in S_n}
\sum_{\beta_1,\beta_2\in S_n}\delta (\beta_1\sigma_1 \beta_2\sigma_1^{-1})
\delta (\beta_1\sigma_2 \beta_2\sigma_2^{-1})\delta (\beta_1\sigma_3 \beta_2\sigma_3^{-1})
\label{Nis3}
\eea
The delta functions appearing in this expression are equal to 1 if the argument of the delta is the identity permutation, and
are otherwise equal to zero.
The first few values of ${\cal N}_3$ are shown in Table \ref{table1} below.
\begin{table}[h]
\centering
\begin{tabular}{|c|c|c|c|c|c|c|}
\hline
 & n=1 & n=2 & n=3 & n=4 & n=5 & n=6 \\
\hline 
$\mathcal{N}_{3}$ & 1 & 4 & 11 & 43 &  161 & 901 \\
\hline
\end{tabular}
\caption{$\mathcal{N}_{3}$ counts the number of gauge invariant operators for rank 3 tensor fields constructed using 
$2n$ fields. We take $N_{i}=\infty$ for $i=1,2,3$, i.e. we have not accounted for any relations between invariants that
appear at finite $N_i$.}\label{table1}
\end{table}

To argue that (\ref{newobs}) are a complete set of $U(N_1)\times U(N_2)$ invariants, we want to reproduce the
counting above, using (\ref{newobs}) as the dynamical variables.
The counting for $n=1,2$ is simple.
For $n=1$ we have a single invariant given by $T(2)^i_i$ in agreement with Table \ref{table1} above.
For $n=2$ we have $T(2)^{i}_{i}T(2)^{j}_{j}$, $T(2)^{i}_{j}T(2)^{j}_{i}$, $T(4)^{ij}_{ij}$ or $T(4)^{ij}_{ji}$.
This gives a total of four gauge invariants, in agreement with Table \ref{table1}.
For higher values of $n$ things are less trivial as there are non-trivial relations between different invariant variables.
To see the origin of these relations, note that $T(2n)$ enjoys a $Z_n$ generated by the $n$ cycle $(12\cdots n)$.
For example, $T(6)$ enjoys a $Z_3$ generated by the three cycle $(123)$.
The symmetry acts as
\bea
T(6)^{i_1i_3i_5}_{i_2i_4i_6}=T(6)^{i_5i_1i_3}_{i_6i_2i_4}=T(6)^{i_3 i_5 i_1}_{i_4 i_6 i_2}
\eea
A nice example in which these symmetries play a role is for $n=4$.
For $n=4$ there are 5 possible combinations of fields that can be used to construct the gauge invariant, namely
\bea
T(2)^{i_1}_{j_1}T(2)^{i_2}_{j_2}T(2)^{i_3}_{j_3}T(2)^{i_4}_{j_4}\quad &{\rm or}& \quad T(4)^{i_1i_2}_{j_1j_2}T(2)^{i_3}_{j_3}T(2)^{i_4}_{j_4}
\quad {\rm or} \quad T(6)^{i_1i_2i_3}_{j_1j_2j_3}T(2)^{i_4}_{j_4}\cr\cr\cr
\quad &{\rm or}& \quad
T(4)^{i_1i_2}_{j_1j_2}T(4)^{i_3i_4}_{j_3j_4}
\quad {\rm or} \quad T(8)^{i_1i_2i_3i_4}_{j_1j_2j_3j_4}
\eea
For each of these the number of gauge invariant variables is given by computing a sum of the form\footnote{This is an
application of the orbit counting lemma. See \cite{deMelloKoch:2011uq} for relevant background.}
\bea
|S_4//G|={1\over |G|}\sum_{h_1\in G}\sum_{g\in S_4}
\delta(h_1 g h_1^{-1} g^{-1})
\eea
The notation used for the LHS of this equation is explained in Appendix \ref{Notation}.
Using an element $g\in S_4$, the invariants constructed from a product of four $T(2)$s are given by 
$T(2)^{i_1}_{i_{g(1)}}T(2)^{i_2}_{i_{g(2)}}T(2)^{i_3}_{i_{g(3)}}T(2)^{i_4}_{i_{g(4)}}$.
There is a symmetry under permuting
the fields so that $G=S_4$ and we find 5 gauge invariant operators. $G$ swaps the $T(2)$s.
For $T(4)^{i_1i_2}_{i_{g(1)}i_{g(2)}}T(2)^{i_3}_{i_{g(3)}}T(2)^{i_4}_{i_{g(4)}}$ we find that $G=S_2\times Z_2$ 
and we find 10 gauge invariant operators. 
$G$ swaps the two $T(2)$s and does cyclic permutations of the upper and lower indices of $T(4)$.
For $T(6)^{i_1i_2i_3}_{i_{g(1)}i_{g(2)}i_{g(3)}}T(2)^{i_4}_{i_{g(4)}}$ we have $G=Z_3$ and we find 10 gauge 
invariant operators. $G$ performs cyclic swaps of the upper and lower indices of $T(6)$.
For $T(4)^{i_1i_2}_{i_{g(1)}i_{g(2)}}T(4)^{i_3i_4}_{i_{g(3)}i_{g(4)}}$ we have $G=S_2\times Z_2\times Z_2$ and 
we find 8 gauge invariant operators. $G$ swaps the two $T(4)$s and performs cyclic permutations of the upper 
and lower indices.
For $T(8)^{i_1i_2i_3i_4}_{i_{g(1)}i_{g(2)}i_{g(3)}i_{g(4)}}$ we have $G=Z_4$ and we find 10 gauge invariant 
operators. $G$ performs cyclic permutation of the upper and lower indices of $T(8)$.
This gives a total of 43 gauge invariant operators which reproduces the $n=4$ entry in Table \ref{table1}.

For $n=6$ we have also verified that there is a total of 901 operators. The pattern follows that of the computations above,
but this example is complicated enough that it is worth the effort to set things up in a more general way. 
To illustrate the comments that follow, consider the number of gauge invariant operators constructed for $n=32$ i.e. we 
use a total of 64 fields. 
Each possible collection of operators that can be used to construct an invariant can be labeled with a Young diagram that 
has $n$ boxes. 
Further, we will see that the group $G$ can be read straight from the Young diagram. 
Each row containing $k$ boxes translates into a tensor $T(2k)^{i_1\cdots}_{j_1\cdots}$.
As an example, the Young diagram
\bea
{\tiny \yng(8,8,4,4,4,2,2)}
\eea
labels the gauge invariant operators constructed using $T(16)T(16)T(8)T(8)T(8)T(4)T(4)$.
To read $G$ from the Young diagram, first note that for each row we get a cyclic group $Z_k$, where $k$
is the number of boxes in the row. For the above Young diagram we'd get
\bea
Z_8 \times Z_8\times  Z_4\times Z_4\times Z_4\times Z_2\times Z_2
\eea
For $p$ rows of equal length we get a factor of $S_p$. For the above Young diagram we get
\bea
S_2\times S_3\times S_2
\eea
$G$ is a product of these cyclic and symmetric groups. For the above Young diagram we'd have
\bea
G=Z_8 \times Z_8\times  Z_4\times Z_4\times Z_4\times Z_2\times Z_2\times S_2\times S_3\times S_2
\eea

To illustrate these rules, we count the number of gauge invariant operators for $n=6$.
For ${\tiny \yng(1,1,1,1,1,1)}$, $G=S_6$ and we find 11 gauge invariant operators.
For ${\tiny \yng(2,1,1,1,1)}$, $G=Z_2\times S_4$ and we find 34 gauge invariant operators.
For ${\tiny \yng(3,1,1,1)}$, $G=Z_3\times S_3$ and we find 58 gauge invariant operators.
For ${\tiny \yng(2,2,1,1)}$, $G=Z_2\times Z_2\times S_2\times S_2$ and we find 70 gauge invariant operators.
For ${\tiny \yng(4,1,1)}$, $G=Z_4\times S_2$ and we find 108 gauge invariant operators.
For ${\tiny \yng(3,2,1)}$, $G=Z_3\times Z_2$ and we find 136 gauge invariant operators.
For ${\tiny \yng(2,2,2)}$, $G=Z_2\times Z_2\times Z_2\times S_3$ and we find 34 gauge invariant operators.
For ${\tiny \yng(5,1)}$, $G=Z_5$ and we find 148 gauge invariant operators.
For ${\tiny \yng(4,2)}$, $G=Z_4\times Z_2$ and we find 108 gauge invariant operators.
For ${\tiny \yng(3,3)}$, $G=Z_3\times Z_3\times S_2$ and we find 58 gauge invariant operators.
For ${\tiny \yng(6)}$, $G=Z_6$ and we find 136 gauge invariant operators.
This gives 901 operators in total, reproducing the $n=6$ entry in Table \ref{table1}.

The counting arguments we have given above provide convincing evidence that the $U(N_1)\times U(N_2)$ invariants
given in (\ref{newobs}) indeed provide a complete list.
The simple model, of rank-3 tensors considered above, is equivalent to a model of interacting $U(N_3)$ tensors of 
every even rank.
We can attempt a similar construction for higher rank tensors again constructing the invariants of two of the
groups appearing.
In this way tensor models of higher rank tensors are described by a collection of tensors that transform in 
all but 2 gauge groups.
Dealing with this remaining gauge symmetry becomes more and more non-trivial as the rank is increased beyond 3
and for this reason we will restrict attention to rank 3 tensors.
In the sections that follow we will consider constructing collective field theory based on a subset of these invariant variables. 

\section{A ``planar limit'' for the tensor model}\label{Strings}

In the previous section we have described the structure of the space of invariants for tensor models.
The complete space is difficult to manage.
Consequently, we seek subsets of invariant variables that are dynamically closed. 
In this section we will find a matrix like subsector, which has non-trivial dynamics but is still simple enough
to manage.

Consider a model with two basic flavors of tensors, $\phi^{abc}$, $\bar\phi_{abc}$ and $ \psi^{abc}$, $\bar\psi_{abc}$.
There are also momenta conjugate to these fields and we have the equal time commutation relations
\bea
[\Pi_{abc},\phi^{def}]&=&-i\delta^d_a\delta^e_b\delta^f_c\qquad\qquad 
[\bar\Pi^{def},\bar\phi_{abc}]=-i\delta^d_a\delta^e_b\delta^f_c\cr
[P_{abc},\psi^{def}]&=&-i\delta^d_a\delta^e_b\delta^f_c\qquad\qquad
[\bar P^{def},\bar\psi_{abc}]=-i\delta^d_a\delta^e_b\delta^f_c
\eea
All commutators between barred and unbarred fields, as well as all others commutators not shown, vanish.
The Hamiltonian for the free theory is
\bea
  H&=&\Pi_{abc}\bar\Pi^{abc}+\omega^2\phi^{abc}\bar\phi_{abc}+P_{abc}\bar P^{abc}
           +\omega^2\psi^{abc}\bar\psi_{abc}\cr\cr
    &=&-{\partial\over\partial\phi^{abc}}{\partial\over\partial\bar\phi_{abc}}+\omega^2 \phi^{abc}\bar\phi_{abc}
-{\partial\over\partial\psi^{abc}}{\partial\over\partial\bar\psi_{abc}}+\omega^2\psi^{abc}\bar\psi_{abc}
\eea
A simple set of invariants that are closed under both the splitting and joining operations of collective field theory, are the 
traces of products of $T_\phi(2)^i_j$ and $T_\psi(2)^i_j$, where
\bea
T_\phi(2)^i_j=\phi^{abi}\bar\phi_{abj}\qquad
T_\psi(2)^i_j=\psi^{abi}\bar\psi_{abj}
\eea
Both transform in the adjoint of $U(N_3)$.
Following \cite{deMelloKoch:2017bvv} we can construct an exact collective description for $T_\phi(2)$.
We will extend the discussion of \cite{deMelloKoch:2017bvv} by considering a model with a quartic potential.
Choosing a potential which is a sum $V(T_\phi(2))+V(T_\psi(2))$ preserves the fact that the dynamics
of $T_\phi(2)$ defines a dynamically closed subsector.
Motivated by this remark, we consider the interacting dynamics of the Hamiltonian
\bea
  H&=&\Pi_{abc}\bar\Pi^{abc}+\omega^2\phi^{abc}\bar\phi_{abc}
           +g\phi^{abc}\bar\phi_{abd}\phi^{efd}\bar\phi_{efc}
\eea
where we dropped the $\psi$ terms since they play no role in the present discussion.
The $\psi$ terms will be needed below when we construct BMN loops.
The collective field
\bea
\phi (x)=\int {dk\over 2\pi}\, e^{-ikx}\, \phi_k\qquad
\phi_k={\rm Tr}(e^{ikT_\phi (2)})
\eea
is the density of eigenvalues of $T_\phi (2)$.
A straight forward computation gives
\bea
\Omega_{k,k'}&=& {\partial\phi_{k}\over\partial \phi^{abc}} {\partial\phi_{k'}\over\partial\bar\phi_{abc}}
=ikk'{\partial\over\partial k}\phi_{k+k'}\cr\cr
\omega_k&=&-{\partial\over\partial \phi^{abc}}\left( {\partial\phi_k\over\partial \bar\phi_{abc}}\right)
k\int_0^1 d\tau\,\, \phi_{\tau k}i{\partial\over\partial \tau}\phi_{(1-\tau )k}
-ikN_1 N_2\phi_{k}
\eea
In position space we obtain\footnote{The formula for $\Omega (x,x')$ coincides with the radial sector of multi matrix 
models and the formula for $\omega (x)$ is very similar - see  
\cite{Masuku:2009qf,Masuku:2011pm,Masuku:2014wxa,Masuku:2015vta}.}
\bea
   \Omega (x,x')&=&{\partial\over\partial x}{\partial\over\partial x'}\left( x\phi (x)\delta (x-x')\right)\cr\cr
\omega (x) &=&2{\partial\over\partial x}\dashint dy\,\, \phi(x)\phi(y){x\over x-y}
+(N_1 N_2-N_3){\partial\phi(x)\over\partial x}
\eea
This easily leads to the Hamiltonian (after discarding constant terms)
\bea
H=\int dx {\partial\pi\over\partial x}x\phi (x){\partial\pi\over\partial x} + V_{\rm eff}
\eea
where the effective potential is
\bea
V_{\rm eff}=\int dx\left[ {\pi^2 x\over 3}\phi^3
+{N_3^2-N_1^2N_2^2\over 4x}\phi (x)+gx^2\phi(x)+\omega^2 x \phi (x)-\mu\phi(x)\right]
\eea
The last term above is a Lagrange multiplier enforcing the constraint $\int dx \phi (x)=N_3$.
The large $N$ classical field minimizes the effective potential, which implies the following equation of motion
\bea
0={\delta V_{\rm eff}\over\delta\phi (x)}=\pi^2 x\phi^2+{N_3^2-N_1^2N_2^2\over 4x}
+\omega^2 x -\mu+gx^2\cr\cr
\Rightarrow \phi(x)={1\over\pi}\sqrt{{\mu\over x}-\omega^2-{N_3^2-N_1^2N_2^2\over 4x^2}-gx\,\,\,\,}
\eea
where $\mu$ is fixed by requiring that
\bea
\int_{0}^{x_0} dx\phi (x)= N_3\qquad
\phi(x_0)=0
\eea
The above dynamics simplifies if we set $N_1 N_2=N_3$, something that we do from now on.
In the Appendix \ref{CollectiveCheck} we demonstrate that this collective field reproduces the correct large $N$ expectation
values in perturbation theory.  

The discussion of this section shows that there is a matrix-like subsector in the Hilbert space of the tensor model.
In particular, there will be string states. 
This has all been extracted from a single matrix subsector.
To see stringy states and their excitations, we will construct the analog of the BMN loops\cite{Berenstein:2002jq}.
We carry out the discussion in the context of the free theory, using both the $\phi^{abc}$ and $\psi^{abc}$ fields.
The fields have the following expansions in terms of creation and annihilation operators
\bea
\phi^{abc}(t)&=&\frac{1}{2\omega}\left( e^{-i\omega t}a^{abc} + e^{i\omega t}b^{abc\,\dagger}\right)
\qquad
\bar\phi_{abc}(t)=\frac{1}{2\omega}\left( e^{-i\omega t}b_{abc} + e^{i\omega t}a_{abc}^{\dagger}\right)\cr
\psi^{abc}(t)&=&\frac{1}{2\omega}\left( e^{-i\omega t}c^{abc} + e^{i\omega t}d^{abc\,\dagger}\right)
\qquad
\bar\psi_{abc}(t)=\frac{1}{2\omega}\left( e^{-i\omega t}d_{abc} + e^{i\omega t}c_{abc}^{\dagger}\right)
\eea
and the oscillators obey the following commutation relations
\bea
[a^{abc},a_{def}^\dagger]=[c^{abc},c_{def}^\dagger]=2\omega\delta^a_d\delta^b_e\delta^c_f
=[b_{def},b^{abc\,\dagger}]=[d_{def},d^{abc\,\dagger}]
\eea
We can also write expansions for the conjugate momenta as follows
\bea
\bar\Pi^{abc}(t)=-\frac{i}{2}\left( e^{-i\omega t}a^{abc} - e^{i\omega t}b^{abc\,\dagger}\right)\qquad
\Pi_{abc}(t)=-\frac{i}{2}\left( e^{-i\omega t}b_{abc} - e^{i\omega t}a_{abc}^{\dagger}\right)\cr
\bar P^{abc}(t)=-\frac{i}{2}\left( e^{-i\omega t}c^{abc} - e^{i\omega t}d^{abc\,\dagger}\right)\qquad
P_{abc}(t)=-\frac{i}{2}\left( e^{-i\omega t}d_{abc} - e^{i\omega t}c_{abc}^{\dagger}\right)
\eea
Expressing the Hamiltonian in terms of oscillators, we have
\bea
   H={1\over 4}\left(a^\dagger_{abc}a^{abc}
                           +b^{abc\,\dagger}b_{abc}
                           +c^\dagger_{abc}c^{abc}
                           +d^{abc\,\dagger}d_{abc}\right) + 2\omega
\eea
Introduce the pair of fields
\bea
Z^i_j\equiv {a^\dagger_{klj}b^{kli\,\dagger}\over 2\omega \sqrt{N_1 N_2}}\qquad Y^i_j\equiv 
{c^\dagger_{klj}d^{\dagger\,kli}\over 2\omega \sqrt{N_1 N_2}}\label{frstdes}
\eea
Using these we can construct the following gauge invariant observables
\bea
\Phi(n_1,n_2,\cdots)={\rm Tr}(Z^{n_1}YZ^{n_2-n_1}YZ^{n_3-n_2}\cdots)
\eea
These loop variables have a large $N$ dynamics that matches the dynamics of BMN loops.
We want to compute correlators of the form
\bea
\langle 0|\Phi(n_1',n_2',\cdots)^\dagger \Phi(n_1,n_2,\cdots)|0\rangle
\eea
Lets start by making a few observations. 
Is is straight forwards to argue that
\bea
\langle 0|(Z^\dagger)^i_j Z^k_l|0\rangle = \delta^k_j\delta^i_l = \langle 0|(Y^\dagger)^i_j Y^k_l|0\rangle
\eea

and

\bea
&&\langle 0|(Z^\dagger)^{i_1}_{j_1} (Z^\dagger)^{i_2}_{j_2} Z^{k_1}_{l_1} Z^{k_2}_{l_2}|0\rangle =
\langle 0|(Y^\dagger)^{i_1}_{j_1} (Y^\dagger)^{i_2}_{j_2} Y^{k_1}_{l_1} Y^{k_2}_{l_2}|0\rangle\cr 
&&=\delta^{i_1}_{l_1}\delta^{k_1}_{j_1}\delta^{i_2}_{l_2}\delta^{k_2}_{j_2}+
\delta^{i_1}_{l_2}\delta^{k_1}_{j_2}\delta^{i_2}_{l_1}\delta^{k_2}_{j_1}+
\frac{1}{N_1 N_2}\left(\delta^{i_1}_{l_1}\delta^{k_1}_{j_2}\delta^{i_2}_{l_2}\delta^{k_2}_{j_1}+
\delta^{i_1}_{l_2}\delta^{k_1}_{j_1}\delta^{i_2}_{l_1}\delta^{k_2}_{j_2}\right)
\eea

The generalization of this formula to $n$ fields is

\bea
\langle 0|(Z^\dagger)^{i_1}_{j_1}\cdots (Z^\dagger)^{i_n}_{j_n} Z^{k_1}_{l_1}\cdots Z^{k_n}_{l_n}|0\rangle &=&
\langle 0|(Y^\dagger)^{i_1}_{j_1}\cdots (Y^\dagger)^{i_n}_{j_n} Y^{k_1}_{l_1}\cdots Y^{k_n}_{l_n}|0\rangle\cr\cr
&=&\sum_{\sigma\in S_n}\sigma^{I}_{L}(\sigma^{-1})^{K}_{J}\left(1+O\left(\frac{1}{N_1 N_2}\right)\right)
\label{SameWick}
\eea
We have introduced a notation \cite{Corley:2001zk} which uses a capital Roman letter to collect the little letter indices, 
so that for example $I$ stands for $i_1,i_2,\cdots,i_n$. 
We refer to $I$ as a multi-index.
The formula (\ref{SameWick}) is a nice result because it is what we would get from 
a matrix model so that we have a detailed and specific
grasp on how the single matrix dynamics emerges from the tensor model.
Note that the above observations continue to work if we use the fields
\bea
Z^i_j\equiv {a^\dagger_{klj}b^{kli\,\dagger}\over 2\omega \sqrt{N_1 N_2}}\qquad\qquad 
Y^i_j\equiv {a^\dagger_{klj}d^{\dagger\,kli}\over 2\omega \sqrt{N_1 N_2}}\label{secdes}
\eea
which employ one less oscillator. 
The subleading corrections to the two descriptions (i.e. (\ref{frstdes}) versus (\ref{secdes})) do not agree, i.e. the
$O\left(\frac{1}{N_1 N_2}\right)$ terms in (\ref{SameWick}) depend on which description one uses.
Again, we stress that the mapping to matrix model correlators holds at the leading order in large $N_1 N_2$.

The BMN spectrum of excited states is obtained by adding an interaction
\bea
\propto {\rm Tr}\left( [Z^\dagger ,Y^\dagger][Z,Y]\right)
\eea
to the free $Z,Y$ matrix model.
It is simple to check that the spectrum of the tensor model quantum mechanics, with loops (\ref{secdes}) and with the interaction
\bea
\propto 
(b^{\dagger mnh}a^\dagger_{qrh}d^{\dagger qri}-d^{\dagger mnh}a^\dagger_{qrh}b^{\dagger qri})
(b_{kli}a^{klj}d_{mnj}-d_{kli}a^{klj}b_{mnj})
\eea
agrees with the spectrum of excited string states, at leading order at large $N_1 N_2$.

The observation (\ref{SameWick}) above has a number of immediate implications.
The subset of tensor model correlators we consider follow by using the usual rules for Wick contracting matrices.
Consequently, we will have many of the matrix model results in this tensor model setting.
This includes the fact that different trace structures don't mix, which implies a Fock space type description of this limit with
the number of traces identified with the number of particles.
At large $N_1 N_2$ gauge invariant observables constructed using $Z$ and $Y$ will have a genus expansion, exactly like the
usual 't Hooft expansion in matrix models.
We will have ribbon graphs, that triangulate surfaces and the genus of these surfaces determines the power of $N_3$ of the
graph.
This is a nice geometrical interpretation for the $1/N_3$ expansion and it provides convincing evidence for emergent
strings\footnote{For another geometrical interpretation of the $N$ dependence see Section 6 of \cite{BenGeloun:2017vwn}.
This interpretation seems to be distinct to the interpretation we obtain here from triangulated worldsheets.}.
We will also have the usual simplifications of large $N$ for matrix models including factorization in the planar limit. 
It is also possible to consider the finite $N_3$ physics of the model, using Schur polynomials \cite{Corley:2001zk}, 
restricted Schur polynomials \cite{Bhattacharyya:2008rb} and other 
methods \cite{Kimura:2007wy,Brown:2007xh,Brown:2008ij} that have been developed for matrix models.

\section{The Collective Field Theory of Melons}\label{Melons}

The space of all possible invariants of the tensor model appears to be  too complicated to manage without further input.
In the previous section we have studied some closed subsectors which are matrix like. 
The dynamics of these subsectors are certainly manageable, but one might worry that they don't exhibit 
behavior which is characteristic of the tensor model.
For that reason, we turn in this section to an approximation scheme in which we take advantage of known results for the
large $N$ limit of tensor models\footnote{Some of the equations we write only hold at the leading order in the large 
$N$ expansion. With this approximation, the collective field theory does not take us outside the space of melonic gauge 
invariants.}.
The novelty of the large $N$ limit of tensor models is that it is dominated by melonic diagrams and it is precisely this
feature that it has in common with SYK.
The large $N$ limit we consider in this section is obtained by setting $N_1=N_2=N_3$.
While reading this section, the reader will find it useful to 
consult \cite{Jevicki:1993rr,Gurau:2011tj,Gurau:2012ix,Bonzom:2012cu,Itoyama:2017wjb,Itoyama:2018but}.
The first result we will make use of is the fact that the large $N$ limit is dominated by melonic 
diagrams\cite{Bonzom:2011zz,Bonzom:2012hw}.
For some examples of melonic diagrams see Figure \ref{MG} below.
\begin{figure}[ht]%
\begin{center}
\includegraphics[width=1.0\columnwidth]{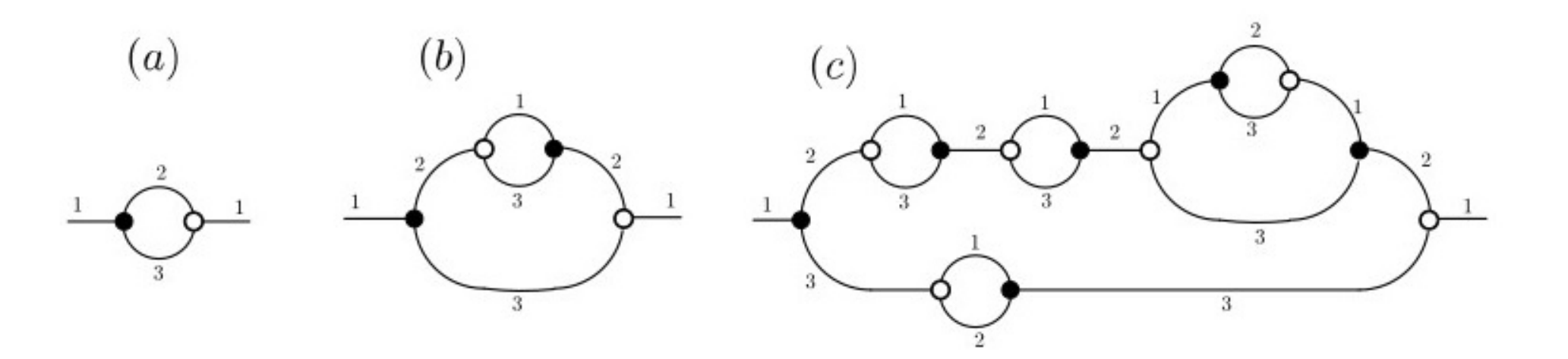}%
\caption{Examples of melonic graphs.}%
\label{MG}%
\end{center}
\end{figure}

By closing the loop in these melonic graphs, we obtain a graphical representation of a gauge invariant operator.
Each white vertex corresponds to a field $\phi^{abc}$ and each black vertex to a field $\bar\phi_{abc}$. 
The lines in the graph are an instruction for how to contract indices to obtain the invariant.
The lines are labeled by an index $i$ telling us that the corresponding index transforms under $U(N_i)$.
We refer to the melon shown in (a) as an elementary melon. 
The melons shown in (b) and (c) are dressed. 
It is clear that each melon, after removing all dressing, is defined by a pair of vertices. 
There is a single Feynman diagram that contributes to the large $N$ limit, given by contracting the pairs of vertices that
define a given melon\footnote{This follows immediately after computing a few examples. Alternatively, consult Section III A
of \cite{Bonzom:2012cu} for a proof that uses the Schwinger-Dyson equations.}.
Later we will see that this gives us important insights into how the collective field theory of melons simplifies.

\begin{figure}[hb]%
\begin{center}
\includegraphics[width=0.9\columnwidth]{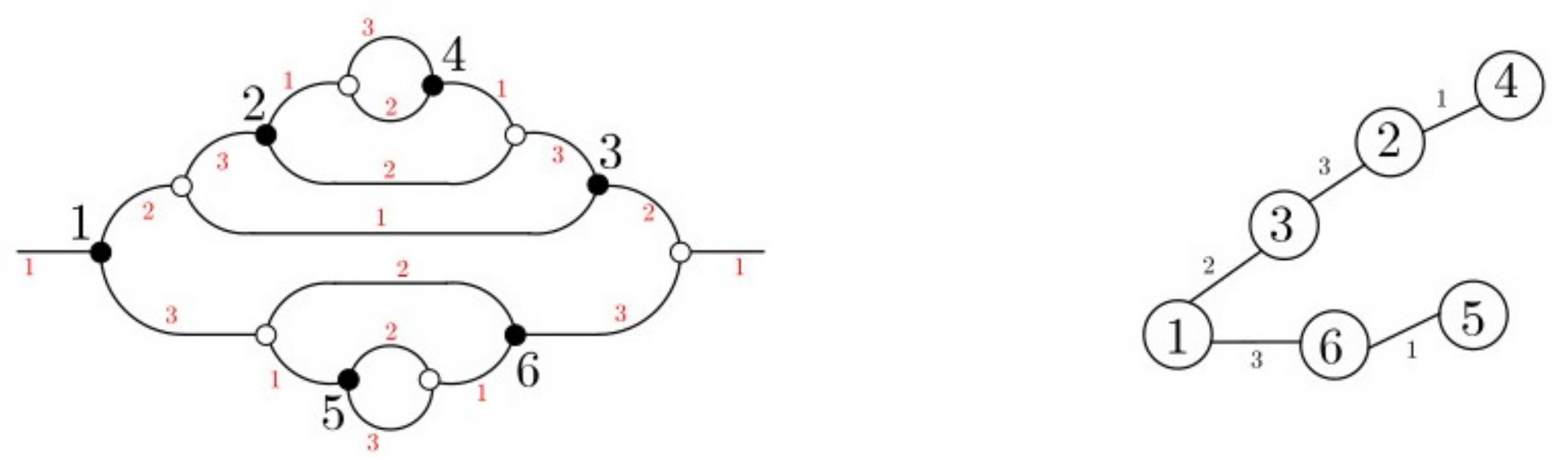}%
\caption{The tree labeling a melonic invariant.}%
\label{tree}
\end{center}
\end{figure}

In what follows, we focus our attention on the invariants formed by closing the loop in the melon graphs.
This is an enormous set of invariant variables. 
However, we can write equations for the resulting dynamics thanks largely to the fact that there is a very nice way to label these 
invariants: they can be labeled by $D$-ary trees, as explained in \cite{Bonzom:2011zz}.
An example of this labeling is given in the Figure \ref{tree} above.
The fact that we can label melonic invariants with trees has immediate implications for collective field theory.
The operation of joining given by $\Omega$ implies that joining melonic invariants keeps us within the space
of melonic invariants.
Denote the loop for a melonic invariant labeled by a tree with $p$ melons $T_p$ by $\phi_{{\cal M}(T_p)}$. 
Joining a tree of $p$-melons and a tree of $p'$-melons gives a tree of $p+p'-1$ melons
\bea
\Omega (\phi_{{\cal M}(T_p)},\phi_{{\cal M}(T_{p'})})=\sum_{T_{p+p'-1}}\phi_{{\cal M}(T_{p+p'-1})}
\eea
The sum on the right hand side above includes $pp'$ terms.
\begin{figure}[ht]%
\begin{center}
\includegraphics[width=0.7\columnwidth]{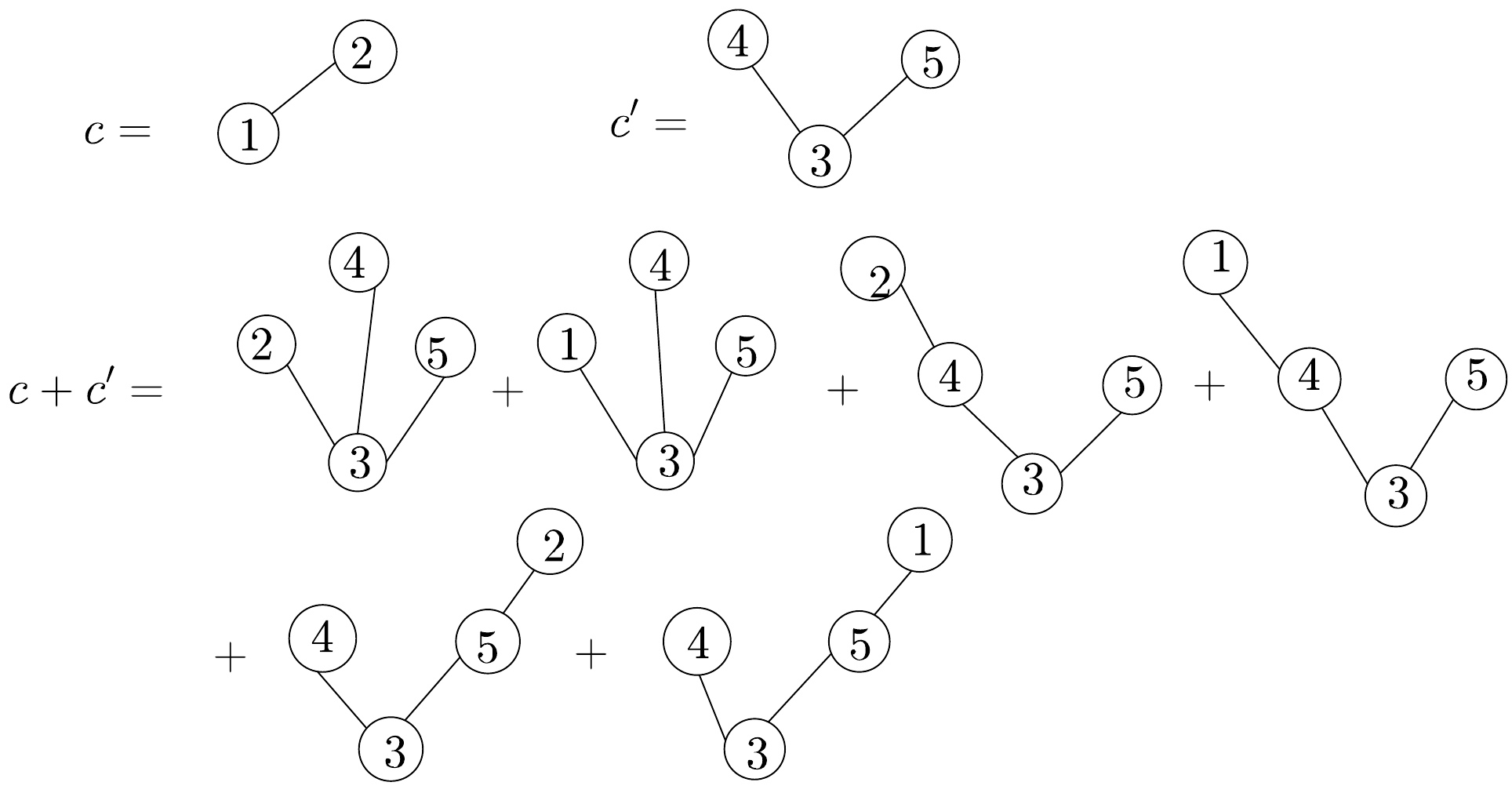}%
\caption{An example of tree joining. This figure shows the possible values for $C+C'$ for the given $C$ and $C'$.
The joining is performed by contracting a solid vertex in $C$ with a hollow vertex in $C'$.}%
\label{join}
\end{center}
\end{figure}

Splitting does not keep us inside the space of melons.
Splitting a single melonic invariant can give us back a melonic invariant (this occurs when we split on vertices
defining an elementary melon), a product of disconnected melonic invariants (this occurs when we split on vertices
defining a melon which is not elementary) or an invariant which is not melonic (this occurs when we split on vertices
that belong to different melons). 
This means that the melonic invariants are not dynamically decoupled and there is mixing between melonic and non-melonic
invariants.
This is a significant complication.
Our approach towards this issue is to treat the problem systematically in the $1/N$ expansion.
At the leading order we can simply omit all terms involving non-melonic invariants, since these are subleading.
Further, the expectation values of disconnected melonic invariants have the same large $N$ behavior as the
original melonic invariant that produced them (because the relevant splitting is a Wick contraction contributing to the unique
Feynman diagram giving the complete large $N$ expectation value of the melonic observable).
The main obstacle to completing the construction of the collective Hamiltonian, is now determining the inverse
of $\Omega (\phi_{{\cal M}(T_p)},\phi_{{\cal M}(T_{p'})})$.
This is a formidable task that is under current investigation.

Although we have not yet managed to determine the collective Hamiltonian, even at large $N$, there is a little more we can say.
To illustrate the argument, consider a Gaussian one matrix model with action
\bea
S=\int dt \left({1\over 2}\Tr \dot{M}^2 -{\omega\over 2}\Tr M^2 \right)
\eea
The collective field is given by
\bea
   \phi (x)=\int {dk\over 2\pi} e^{-ikx}\phi_k \qquad \phi_k=\Tr (e^{ikM})
\eea
Summing the planar diagrams we find
\bea
   \langle\phi_k\rangle = \sum_{l=0}^\infty {(ik)^{2l}\over (2l)!}\langle \Tr (M^{2l})\rangle
=\sum_{l=0}^\infty {(ik)^{2l}\over (2l)!}{N^{l+1}\over\omega^l}{(2l-1)!!\over (l+1)!}
={\sqrt{2N\omega}\over k}J_1\left(\sqrt{2N\over\omega}k\right)
\eea 
where $J_1(\cdot)$ is a Bessel function of the first kind.
After a Fourier transform we obtain the usual Wigner distribution
\bea
\langle\phi(x)\rangle = {\omega\over\pi}\sqrt{{2N\over\omega}-x^2}
\eea
describing the large $N$ eigenvalue density.
The coordinate $x$ here is the extra holographic coordinate that emerges from collective field theory.
We will now carry out the corresponding computation in the free tensor model.
Denote $N_1=N_2=N_3=N$.
Denote the melonic variable (before closing the loop as in Figure \ref{MG}) by ${\cal M}(T_p)$.
As a consequence of the fact that a single Feynman diagram contributes, we find
\bea
  \langle \Tr  ({\cal M}(T_p)^n)={N^{1+2pn}\over (2\omega)^{pn}}
\eea
Defining the collective field
\bea
   \phi (x)=\int {dk\over 2\pi} e^{-ikx}\phi_k \qquad \phi_k=\Tr (e^{ik{\cal M}(T_p)})
\eea
and working as we did above, we find
\bea
\langle\phi(x)\rangle = N\delta \left( x-{N^{2p}\over(2\omega)^p}\right)
\eea
The density has support at a single point so that no holographic dimension emerges.
This is a consequence of the simplicity of the melonic limit, namely that a single Feynman diagram contributes.

\section{Finite $N$ Contributions}\label{Expansions}

An interesting limit of matrix model dynamics, quite distinct from the planar limit, is the limit of finite $N$.
This limit of the matrix model is accessed by studying operators built from a parametrically (in $N$) large
number of fields \cite{Balasubramanian:2001nh,Aharony:2002nd,Berenstein:2003ah}.
While the planar limit is dual to the fundamental string and its excitations, finite $N$ sectors of the theory
are dual to giant graviton branes and new space time geometries, including black 
holes \cite{McGreevy:2000cw,Corley:2001zk,Lin:2004nb}.
They also provide an interesting window into the structure of the large $N$ expansion: in this limit one
finds non-perturbative contributions that obstruct the perturbative $1/N$ expansion \cite{deMelloKoch:2019dda}.
Another important point that should be stressed is that these non-perturbative effects are already visible
in the free matrix model since they are concerned with the expansion in $1/N$ and not with the 't Hooft coupling expansion.
Given this motivation, it is clearly interesting to probe the analogous limits of free tensor models, something that is
within reach of existing methods.

One approach towards the finite $N$ physics of the tensor model is to use representation theory techniques, as developed
in \cite{Geloun:2013kta,deMelloKoch:2017bvv}.
The basic idea is to construct gauge invariant operators as a product of the basic fields multiplied by a projection operator. 
Correlation functions reduce to traces of products of projectors.
This implies that they can be computed exactly in the free theory.
The projectors are labeled by Young diagrams and finite $N$ effects are encoded as a cut off on the number of rows 
in the Young diagram labeling the operator.
In what follows we will simply state and use the results we need.
The reader requiring more details is encouraged to consult \cite{Geloun:2013kta,deMelloKoch:2017bvv}, as well as
\cite{Diaz:2017kub,Diaz:2018xzt,Diaz:2018zbg,Mironov:2017aqv}.

We consider products of tensors to build the general gauge invariant operator.
As an example, we could use
\bea
\phi^{i_1j_1k_1}\phi^{i_2j_2k_2}\cdots\phi^{i_nj_nk_n}
\eea
In terms of multi-indices (see the discussion after (\ref{SameWick})) we can write
\bea
\Phi^{IJK}=\phi^{i_1j_1k_1}\phi^{i_2j_2k_2}\cdots\phi^{i_nj_nk_n}\qquad\qquad
\bar\Phi_{IJK}=\bar\phi_{i_1j_1k_1}\bar\phi_{i_2j_2k_2}\cdots\bar\phi_{i_nj_nk_n}
\eea
The basis we study is as follows
\bea
{\cal O}_{r_1,r_2,r_3}^{\gamma_1\gamma_2}=
\sum_{\sigma_1\in S_n}\sum_{\sigma_2\in S_n}\sum_{\sigma_3\in S_n}
C_{r_1,r_2,r_3}^{\gamma_1\gamma_2}(\sigma_1,\sigma_2,\sigma_3)
{\cal O}(\sigma_1,\sigma_2,\sigma_3)\label{btrsp}
\eea
where
\bea
{\cal O}(\sigma_1,\sigma_2,\sigma_3)=\bar\Phi_{IJK}\Phi^{\sigma_1(I)\sigma_2(J)\sigma_3(K)}
\eea
\bea
C_{r_1,r_2,r_3}^{\gamma_1\gamma_2}(\sigma_1,\sigma_2,\sigma_3)=
B^{\gamma_1}_{\alpha_1\alpha_2\alpha_3}
\Gamma^{r_1}{}_{\alpha_1\beta_1}(\sigma_1)
\Gamma^{r_2}{}_{\alpha_2\beta_2}(\sigma_2)
\Gamma^{r_3}{}_{\alpha_3\beta_3}(\sigma_3)
B^{\gamma_2}_{\beta_1\beta_2\beta_3}
\eea
is in fact a restricted character, in the language introduced in \cite{deMelloKoch:2007nbd,Pasukonis:2013ts}.
Thus, (\ref{btrsp}) provides the restricted Schur polynomial basis for the gauge invariant operators of the bosonic
tensor model.
In this formula, $\Gamma^{r}{}_{\alpha\beta}(\sigma)$ denotes the matrix (with row label $\alpha$ and column label
$\beta$) representing $\sigma\in S_n$ in irreducible representation $r$, and we have made use of the branching
coefficients $B^\gamma_{\alpha_1\alpha_2\alpha_3}$ defined by
\bea
   {1\over n!}\sum_{\sigma\in S_n}
\Gamma^{r_1}{}_{\alpha_1\beta_1}(\sigma)
\Gamma^{r_2}{}_{\alpha_2\beta_2}(\sigma)
\Gamma^{r_3}{}_{\alpha_3\beta_3}(\sigma)
=\sum_\gamma B^\gamma_{\alpha_1\alpha_2\alpha_3}B^\gamma_{\beta_1\beta_2\beta_3}
\eea
The branching coefficients provide an orthonormal basis for the subspace of $r_1\otimes r_2\otimes r_3$ that
carries the trivial representation, i.e.
\bea
B^{\gamma_1}_{\alpha_1\alpha_2\alpha_3}B^{\gamma_2}_{\alpha_1\alpha_2\alpha_3}
=\delta^{\gamma_1\gamma_2}
\eea
and where we employ the usual convention that repeated indices are summed.
The advantage of the restricted Schur polynomial basis follows because we are able, in the free theory, to compute
correlators exactly.
The results we will use are \cite{deMelloKoch:2017bvv}
\bea
\langle {\cal O}^{\gamma_1\gamma_2}_{r_1 r_2 r_3}\rangle
&=&n! f_{r_1}(N_1)f_{r_2}(N_2)f_{r_3}(N_3)\delta^{\gamma_1\gamma_2}\label{bos1pnt}
\eea
\bea
\langle :{\cal O}^{\gamma_1\gamma_2}_{r_1 r_2 r_3}:\,
:{\cal O}^{\gamma_3\gamma_4}_{s_1 s_2 s_3}: \rangle
=(n!)^2\delta_{r_1s_1}\delta_{r_2s_2}\delta_{r_3s_3}f_{r_1}(N_1)f_{r_2}(N_2)f_{r_3}(N_3)
{n!\over d_{r_1}}{n!\over d_{r_2}}{n!\over d_{r_3}}\delta^{\gamma_1\gamma_4}\delta^{\gamma_2\gamma_3}\cr
\label{bos2pnt}
\eea
If we normalize the operators to have unit one point function as $N\to\infty$
\bea
    O^{\gamma_1\gamma_2}_{r_1 r_2 r_3}=
{{\cal O}^{\gamma_1\gamma_2}_{r_1 r_2 r_3}\over
n! f_{r_1}(N_1)f_{r_2}(N_2)f_{r_3}(N_3)}
\eea
the two point function becomes
\bea
\langle :O^{\gamma_1\gamma_2}_{r_1 r_2 r_3}:\,
:O^{\gamma_3\gamma_4}_{s_1 s_2 s_3}: \rangle
=\delta_{r_1s_1}\delta_{r_2s_2}\delta_{r_3s_3}{1\over f_{r_1}(N_1)f_{r_2}(N_2)f_{r_3}(N_3)}
{n!\over d_{r_1}}{n!\over d_{r_2}}{n!\over d_{r_3}}\delta^{\gamma_1\gamma_4}\delta^{\gamma_2\gamma_3}\cr
\eea
Now, for simplicity, set $N_1=N_2=N_3=N$ and consider the case that $r_1=r_2=r_3=1^N$, that is, $r$ is a single column
with $N$ boxes.
This is the melonic large $N$ limit used in Section \ref{Melons} and not the matrix-like large $N$ limit of Section \ref{Strings}. 
In the matrix model case this choise of representations 
corresponds to a giant graviton and we would expect this computation to exhibit
non-perturbative effects, even for a real matrix where the one point function doesn't vanish.
The giant graviton description nicely explains the stringy exclusion principle as a bound that exists because the
giant graviton brane stretches to its maximum size in a compact space - see \cite{McGreevy:2000cw} for a detailed discussion.
We have explicitly verified that the corresponding matrix model correlators receive non-perturbative corrections 
in Appendix \ref{RealGiant}.
The tensor model two point function becomes
\bea
\langle :O^{\gamma_1\gamma_2}_{1^N 1^N 1^N}:\,
:O^{\gamma_3\gamma_4}_{1^N 1^N 1^N}: \rangle
=\delta_{r_1s_1}\delta_{r_2s_2}\delta_{r_3s_3}\delta^{\gamma_1\gamma_4}\delta^{\gamma_2\gamma_3}
\eea
This is an exact result, with no sign of any non-perturbative effects, and no hint into the mechanism behind the
finite $N$ cut off.
This is a significant difference as compared to matrix model physics.

Next consider the case that we set $N_1=N_2=N_3$ (i.e. melonic large $N$) and consider the case that $r_1=r_2=r_3=N$, 
that is, $r$ is a single row with $N$ boxes.
This choice of $r_1,r_2,r_3$ would correspond, in the matrix model description, to dual giant gravitons.
In this case
\bea
\langle :O^{\gamma_1\gamma_2}_{1^N 1^N 1^N}:\,
:O^{\gamma_3\gamma_4}_{1^N 1^N 1^N}: \rangle
=\delta_{r_1s_1}\delta_{r_2s_2}\delta_{r_3s_3}\sqrt{\pi^3 N^3}e^{6 N \log (2N)-6N}
\delta^{\gamma_1\gamma_4}\delta^{\gamma_2\gamma_3}
\eea
This is only the leading order result. It is clearly signaling non-perturbative corrections to the large $N$ expansion, 
which is similar to what we find for the matrix model physics. 

\section{Conclusions}\label{Conclusions}

In this article our goal has been to explore the holography of tensor models using collective field theory.
We have obtained a number of definite results that we will summarize in this section.
The first step in a collective field theory description entails identifying the space of gauge invariant variables.
We have developed an approach towards describing the space of gauge invariant variables associated to tensor models,
which is complicated and rich. Our first result is that
\begin{itemize}
\item
For a rank 3 tensor model with gauge group $U(N_1)\times U(N_2)\times U(N_3)$ , we have shown that the gauge
invariant variables for the $U(N_1)\times U(N_2)$ gauge symmetries are tensors of the $U(N_3)$ gauge group.
There is a single tensor with $k$ indices transforming in the fundamental and $k$ indices transforming in the anti 
fundamental for $k=1,2,\cdots$.
\end{itemize}
There is an enormous space of gauge invariants that can be constructed from this set of $U(N_3)$ tensors.
Through simple counting arguments we have given non-trivial evidence that the resulting set of gauge invariants is complete.

We have expressed the dynamics of the tensor model in terms of the invariant variables.
Our initial strategy was to identify sets of invariant variables that are dynamically closed.
\begin{itemize}
\item
Closed subsets that mimic the dynamics of a matrix model have been identified in Section \ref{Strings}, and they define 
a non-linear large $N$ theory that is very similar to the large $N$ dynamics of a matrix model.
The interaction of the effective matrix degrees of freedom is attractive causing eigenvalues to clump.
In moving from the matrix to gauge invariant degrees of freedom we effectively generate a Van der Monde like repulsion. 
The net result of the competition of these two effects is that there is a non-trivial function describing the density
of eigenvalues and this leads to an extra holographic dimension.
\end{itemize}
This is precisely the same mechanism at work in the $c=1$ string \cite{Das:1990kaa}, which is the only example
in which a holographic duality has been proved.
We stress that we can obtain this limit by holding ${N_1N_2\over N_3}$ fixed and taking $N_3\to\infty$, 
which is quite different from the large $N$ limit that is dominated by melons.
By working with a second tensor field, we have defined complex combinations of fields that transform in the
adjoint of $U(N_3)$.
\begin{itemize}
\item
Wick contraction of these complex composite fields are identical, at large $N_1N_2$, to the Wick contraction of matrix fields.
This has allowed us to recover a ribbon graph expansion, factorization and BMN loops for this sector of the theory.
\end{itemize}

In a distinct effort to develop dynamics characteristic of the tensor model, we have explored the large $N$ melonic
limits of the theory.
This limit fixes ${N_1\over N_2}$ and ${N_1\over N_3}$ and takes $N_3\to\infty$.
The resulting collective field theory has a number of promising features: the joining operation leaves 
one within the space of melonic invariants.
Although splitting does not leave us in the space of invariants, excursions away from melonic gauge invariants is
suppressed at large $N$. 
Motivated by these initial encouraging signs, we have considered the collective field theory of melons.
Although we have not managed to write a closed form for the resulting collective field theory Hamiltonian, we have
carried out a direct evaluation of the large $N$ collective field.
\begin{itemize}
\item
We find that the collective field relevant for the melonic limit has delta function support and 
that an extra holographic dimension does not emerge.
\end{itemize}
This is a consequence of the fact that a single Feynman diagram determines the large $N$ value of melonic gauge invariants,
so that the Schwinger-Dyson equations become linear, as commented in \cite{Bonzom:2012cu}.
The non-linearity of the Van der Monde term is suppressed in the melonic limit.

\begin{figure}[ht]%
\begin{center}
\includegraphics[width=0.8\columnwidth]{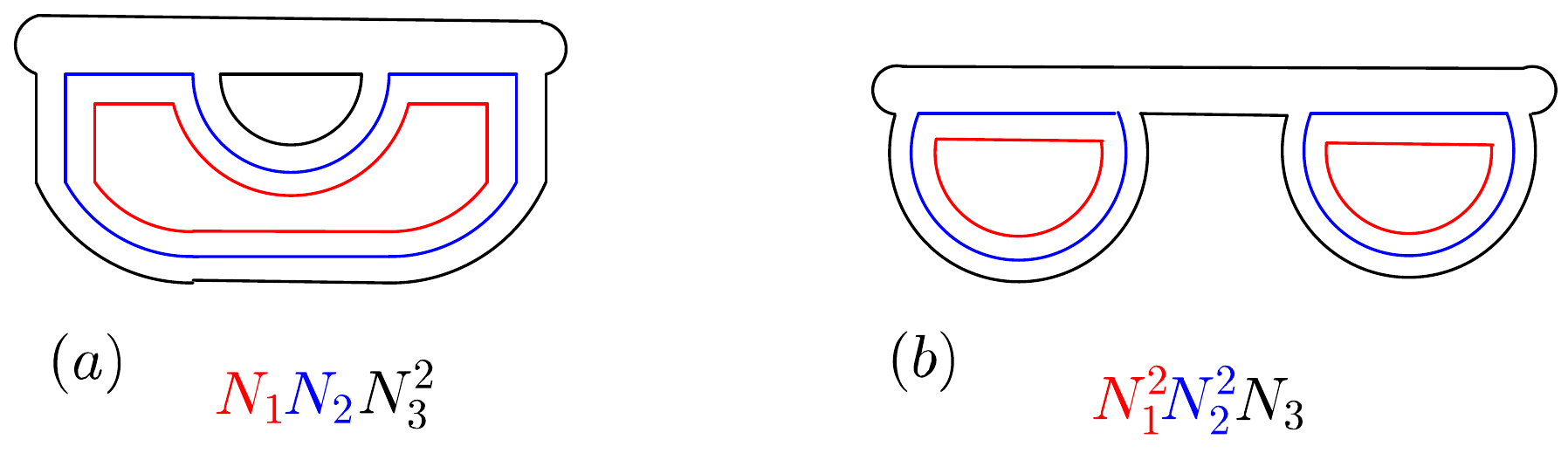}%
\caption{The Feynman diagrams that contribute to $\langle \phi^{abc}\bar{\phi}_{abd}\phi^{efd}\bar{\phi}_{efc}\rangle$.}
\label{puzzle}
\end{center}
\end{figure}

The fact that there is an extra holographic dimension emerging in the matrix like limit that we have considered but
not in the melonic limit, is a consequence of how we take the large $N$ limit in these different settings.
In Figure \ref{puzzle} we have shown the Feynman diagrams that contribute to 
$\langle \phi^{abc}\bar{\phi}_{abd}\phi^{efd}\bar{\phi}_{efc}\rangle$. In our matrix model limit we set $N_1 N_2=N_3$
so that both diagrams survive, matching the fact that two planar diagrams contribute to the large $N$ limit of
$\langle {\rm Tr}(Z\bar ZZ\bar Z)\rangle$. 
On the other hand, in the melonic large $N$ limit, where we set $N_1=N_2=N_3$, diagram (a) is suppressed and the large
$N$ value is completely given by diagram (b).

One of the claims often made in the literature, is that tensor models and tensor field theories admit a $1/N$ expansion 
and a melonic large $N$ limit which is simpler than the planar limit of random matrices and richer than the 
large $N$ limit of vector models.
This gives hope that tensor models may provide an interesting toy model that sheds light on the large $N$ limit of matrix models.
Our results imply that the melonic limit does not lead to a loop space that has enough structure to provide a useful insight 
into how to organize the loop space of the matrix model. 
In contrast to the matrix model, there is no holographic dimension emerging from the color combinatorics.
To see the emergence of an extra holographic dimension, in the melonic large $N$
limit, it seems to be necessary to look at a bilocal description, along the lines of similar discussions for 
SYK\cite{Jevicki:2016bwu,Jevicki:2016ito,Das:2017pif}.
In this collective field theory description, one changes to bilocal (two time) variables which is accompanied by a non-trivial
Jacobian \cite{deMelloKoch:1996mj} which generates non-linear interactions.
This is a key message of our study: for holography in the melonic limit, one must use a bilocal description since
an extra holographic dimension does not emerge from the combinatorics of the gauge indices.

We have considered non-perturbative contributions to the $1/N$ expansion, in the large $N$ melonic limit\footnote{For
the matrix like large $N$ limit we have verified the answer matches the matrix model dynamics - the analysis contains 
no surprises.}.
\begin{itemize}
\item
We have identified non-perturbative contributions to the $1/N$ expansion of invariant observables
in the tensor model.
\end{itemize} 
It would be interesting to find the physical interpretation of these contributions. 

We have initiated a study of the collective field theory of the melonic limit.
Completing this description would provide a detailed understanding of corrections to the large $N$ melonic limit.
It is evident from the description we have sketched that this leads to a theory of interacting random trees
which appears to make contact with the ideas put forward in \cite{Delporte:2019tof}.

{\vskip 0.1cm}

\noindent
\begin{centerline} 
{\bf Acknowledgements}
\end{centerline} 

We would like to thank Sanjaye Ramgoolam and Jo\~ao Rodrigues for useful discussions.
This work is supported by the South African Research Chairs Initiative of the Department of Science and Technology and 
National Research Foundation of South Africa as well as by funds received from the National Institute for Theoretical 
Physics (NITheP).

\begin{appendix}

\section{Notation for Equivalence Classes}\label{Notation}

In Section \ref{Invariants}, we have counted the number of invariants, by taking into account certain symmetries
of the variables involved. 
This amounts to formulating an equivalence relation which implements the symmetry and then counting the resulting
number of equivalence classes.
Equivalences classes associated to the equivalence $g_1 \sim g_2$ if $g_2 = h g_1$ for $g_1, g_2 \in G$ and $h \in H$
lead to the usual notion of a coset. 
The equivalence just described is the right coset $H \setminus G$.
The equivalence $g_2\sim g_1$ if $g_2 = g_1 h$ defines elements of the left coset $G/H$. 

The equivalences classes used in Section \ref{Invariants} are of a different type. 
They can be related to our usual notion of conjugacy classes.
The equivalence relation used to define conjugacy classes says $g_1 \sim g_2$ if $g_1 = g g_2 g^{-1}$ 
for some $g \in G$. 
Although the conjugacy classes are not related to cosets, they are related to double cosets.
The conjugacy classes are given by the double coset
\bea
{\rm Diag} ( G ) \setminus  ( G \times G  ) / {\rm Diag} ( G )
\eea
Here Diag denotes the diagonal action of $G$ obtained by allowing $G$ to act in the same way on both factors in $G\times G$.    
The equivalence is 
\bea
 (g_1,g_2) \sim (g_3g_1g_4,g_3g_2g_4)
\eea 
Now, choose $g_3 = g_1^{-1}$ to map $(g_1,g_2)$ to $(1,g_1^{-1}g_2)$. Then the $g_4$ action leads to
\bea 
   ( 1 , g_4^{-1} ( g_1^{-1} g_2) g_4 ) 
\eea

We will use the notation $G//H$ when we want to talk about the equivalence classes of the relation $g_1 \sim g_2$ 
if $g_1 = h g_2 h^{-1}$ for $g_1,g_2\in G$ and for some $h \in H$.

\section{Perturbative Collective Field Theory}\label{CollectiveCheck}

\begin{figure}[ht]%
\begin{center}
\includegraphics[width=1.0\columnwidth]{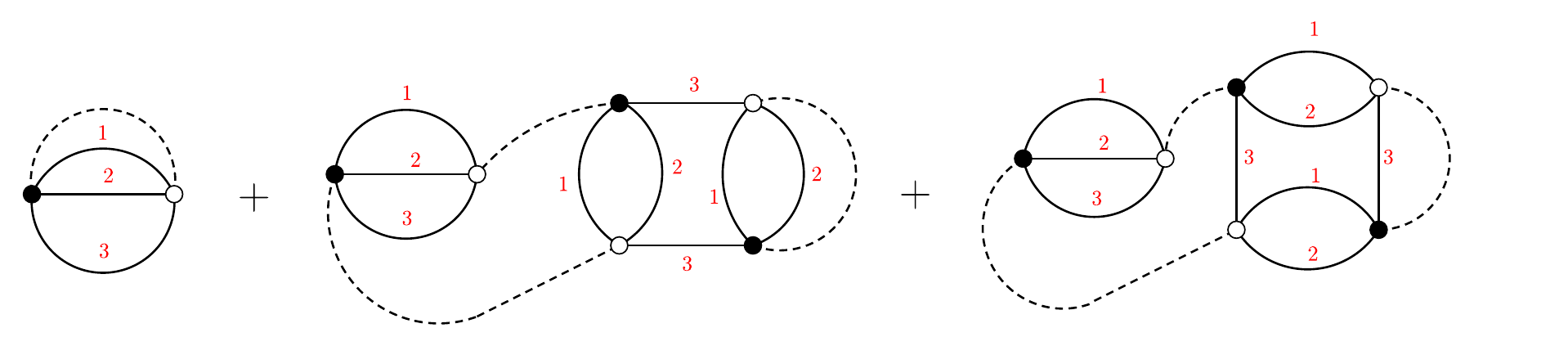}%
\caption{The Feynman diagrams contributing to $\langle\phi^{abc}\bar\phi_{abc}\rangle$. The dashed lines indicate
which operators are Wick contracted.}%
\label{Fig5}%
\end{center}
\end{figure}

In this section we set $N_1N_2=N_3$, which simplifies a number of the formulas that follow.
We also set the frequency $\omega={1\over 2}$.
The collective field is given by
\bea
\phi (x)={1\over\pi}\sqrt{{\mu\over x}-{1\over 4}-gx}
\eea
where the chemical potential $\mu$ is fixed by the requirement
\bea
\int^{x_0}_0\phi (x) dx = N_3
\eea
where $x_{0}$ is the positive zero of the equation
\bea
x - 4\mu +4gx^2 = 0
\eea
At weak coupling $g$, we find
\bea
  x_0={-1+\sqrt{1+64\mu g}\over 8g}=4\mu -64\mu^2 g+O(g^2)
\eea
It is now straight forward to find
$\mu=N_3+6gN_3^2+O(g^2).$
%
Using this collective field we can compute correlators, at large $N$, perturbatively.
As an example
\bea
\langle \phi^{abc}\bar\phi_{abc}\rangle &=& N_3^3 G(t,t)- 4ig N_3^3
\int_{-\infty}^\infty dt' G(t,t')^2 G(t',t') +O(g^2)\cr\cr
&=&N_3^2-8 g N_3^3+O(g^2)
\eea
where we have used the Green's function
\bea
G(t_1,t_2)=\int_{-\infty}^\infty {dE\over 2\pi}{i\over E^2-{1\over 4}+i\epsilon}e^{iE(t_1-t_2)}
\eea
The diagrams summed to obtain this result are show in Figure \ref{Fig5}.
The collective field computation reproduces this result
\bea
\int_0^{x_+}\phi(x) x dx\,=\,N_3^2-8 g N_3^3+O\left(g^2\right)
\eea

\section{Relating Matrix Model and Tensor Model Correlators}

\begin{figure}[ht]%
\begin{center}
\includegraphics[width=0.7\columnwidth]{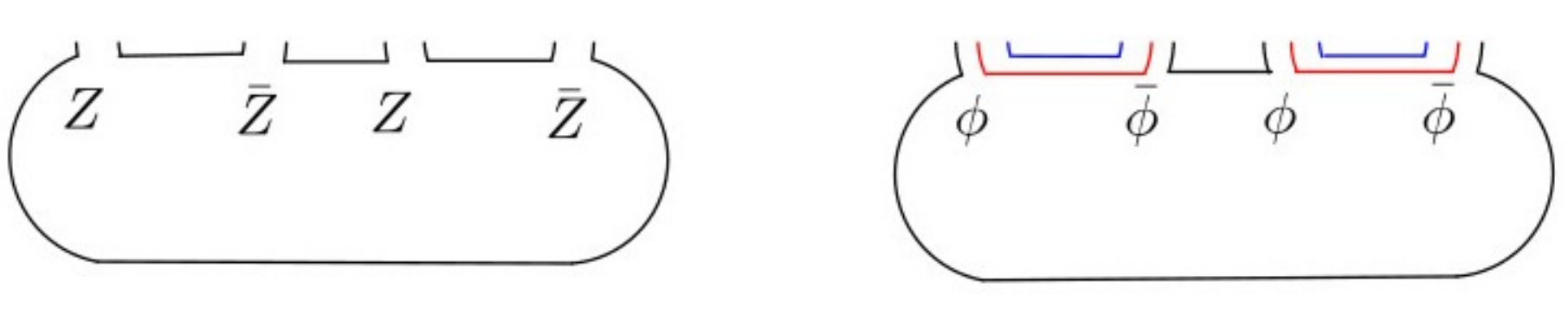}%
\caption{$\Tr (Z\bar ZZ\bar Z)$ is shown on the left and $\phi_{jka}\bar\phi^{jkb}\phi_{lmb}\bar\phi^{lma}$ on the right.}
\label{OpsDef}%
\end{center}
\end{figure}

In Section \ref{Strings} we have argued that there is a subsector of the tensor model that is closely related to a matrix
model. In this Appendix we exhibit a connection between the correlators of the two. 
Consider first correlators in the tensor model.
Assuming the basic contraction
\bea
\langle \phi_{ajk}\bar\phi^{blm}\rangle=\delta^b_a\delta^l_j\delta^m_k\label{eqId1}
\eea
we easily find 
\bea
 \langle \phi_{jka}\bar\phi^{jkb}\phi_{lmb}\bar\phi^{lma}\rangle
 &=& N_1^2 N_2^2N_3 + N_1 N_2 N_3^2  \cr\cr
\langle \phi_{jka}\bar\phi^{jkb}\phi_{lmb}\bar\phi^{lmc}\phi_{pqc}\bar\phi^{pqa}\rangle &=&
3N_1^2N_2^2N_3^2 + N_1N_2N_3^3 + N_1^3N_2^3 N_3 + N_1N_2N_3\cr\cr
\langle\phi_{jka}\bar\phi^{jkb} \phi_{lmb}\bar\phi^{lmc}\phi_{pqc}\bar\phi^{pqd}\phi_{rsd}\bar\phi^{rsa}\rangle
&=& N_1^4N^4_2N_3 + 5N_1^2N^2_2N_3 + 5N_1N_2N_3^2\cr\cr
&~&\qquad + 6N^3_1N^3_2N^2_3 + 6N^2_1N^2_2N^3_3 + N_1N_2N_3^4\cr
&&
\eea
Now consider correlators in the matrix model.
Assuming the basic contraction
\bea
\langle Z^i_j \bar Z^k_l\rangle =\delta^i_l\delta^k_j
\eea
we easily find
\bea
\langle {\rm Tr}(Z\bar ZZ\bar Z)\rangle &=& 2N^3\cr\cr
\langle {\rm Tr}(Z\bar ZZ\bar ZZ\bar Z)\rangle &=& 5N^4 + N^2\cr\cr
\langle {\rm Tr}(Z\bar ZZ\bar ZZ\bar ZZ\bar Z)\rangle &=& 10N^3 + 14N^5
\eea
Notice that if we set $N_1 N_2=N$ and $N_3=N$, the tensor model correlators map into the matrix model correlators.
Its simple to understand this rule diagrammatically: the correspondence holds diagram by diagram.
Consider the operators shown in Figure \ref{OpsDef}

\begin{figure}[ht]%
\begin{center}
\includegraphics[width=0.7\columnwidth]{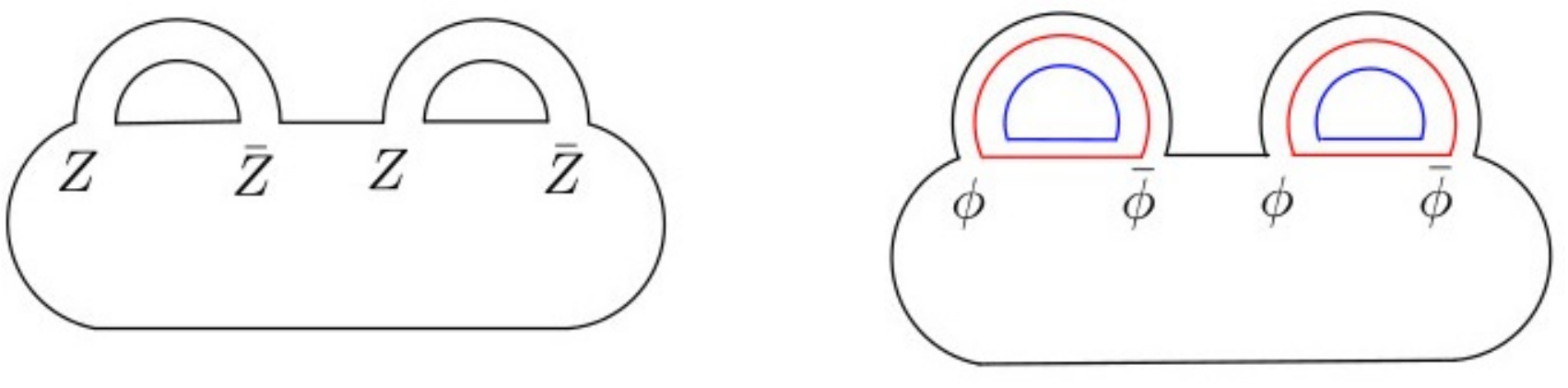}%
\caption{A Feynman diagram contributing to $\langle \Tr (Z\bar ZZ\bar Z)\rangle$ is shown on the left and to
$\langle\phi_{jka}\bar\phi^{jkb}\phi_{lmb}\bar\phi^{lma}\rangle$ on the right. }%
\label{FDiaG}%
\end{center}
\end{figure}

The black line in the tensor model operator is associated to $U(N_3)$, the red line to $U(N_1)$ and the blue line to $U(N_2)$.
One Feynman diagram contributing to the correlator is shown in Figure \ref{FDiaG}. 
Notice that in the tensor model graph, the red and blue lines follow each
other, and if they are collapsed into a single black loop, we recover the matrix model diagram. 
This is the explanation for the rule we have found.

There are a number of interesting results implied by this correspondence between the correlators. 
For example, factorization in the matrix model immediately implies that correlators of gauge invariant observables in the
tensor model factorize at the leading order at large $N$.
The fact that the correspondence holds diagram by diagram immediately implies that the Feynman diagrams contributing to
tensor model observables can be identified as triangulations of string world sheets, with the $N$ dependence of the graph
fixed by the topology of the worldsheet.   

\section{Some Matrix Model Correlators}\label{RealGiant}

In this section we will consider correlators of Schur polynomials of a free Hermittian matrix model.
As far as we are aware, these results are new.
They are obtained using ideas and formulas developed in \cite{Koch:2010zza}.
The Schur polynomial is given by
\bea
\langle\chi_R(M)\rangle
={1\over (2n)!}\sum_{\sigma\in S_{2n}}\chi_R(\sigma){\rm Tr}(\sigma M^{\otimes 2n})
\eea
where $R\vdash 2n$, i.e. $R$ is a Young diagram with $2n$ boxes.
Wick's theorem for the hermittian matrix can be expressed as 
\bea
\langle M^{i_1}_{j_1}M^{i_2}_{j_2}\cdots M^{i_{2n}}_{j_{2n}}\rangle
=\sum_{\sigma\in [2^n]}\sigma^I_J
\eea
Using this result, we find that
\bea
\langle\chi_R(M)\rangle
&=&{1\over (2n)!}\sum_{\sigma\in S_{2n}}\chi_R(\sigma)\langle {\rm Tr}(\sigma M^{\otimes 2n})\rangle \cr
&=&{1\over (2n)!}\sum_{\sigma\in S_{2n}}\sum_{\psi\in [2^n]}\chi_R(\sigma){\rm Tr}(\sigma \psi)\cr
&=&{1\over (2n)!}\sum_{\sigma\in S_{2n}}\sum_{\psi\in [2^n]}\sum_{T\vdash 2n}
\chi_R(\sigma)\chi_T(\sigma \psi) {\rm Dim}_T\cr
&=&{1\over (2n)!}\sum_{\sigma\in S_{2n}}\chi_R(\sigma)f_R
\eea
For the case that $R$ is a single column with $2n=N$ this becomes
\bea
\langle\chi_R(M)\rangle = \langle\chi_{(1^N)}(M)\rangle = (-1)^n (2n-1)!!
\eea
For the case that $R$ is a single row with $2n=N$ this becomes
\bea
\langle\chi_R(M)\rangle = \langle \chi_{(N)}(M) = (2n-1)!! {(4n-1)!\over (2n)!(2n-1)!}
\eea
The two point function of normal ordered operators is
\bea
\langle :\chi_R(M):\,:\chi_S(M):\rangle =\delta_{RS}f_R
\eea
We now introduce the normalized operators
\bea
O_{(1^N)}\equiv {\chi_{(1^N)}\over (N-1)!!}\qquad
O_{(N)}\equiv {N!(N-1)!\over (N-1)!! (2N-1)!}\chi_{(N)}
\eea
The two point functions of these correlators are 
\bea
\langle :O_{(1^N)}:\, :O_{(1^N)}:\rangle &=&\sqrt{N\pi\over 2}\cr
\langle :O_{(N)}:\, :O_{(N)}:\rangle &=&\sqrt{2} N\pi e^{-N {\rm log}4}
\eea
Notice that both of these results are non-perturbative in $N$.
As explained in Section \ref{Expansions}, the analog of these correlators in the tensor model have a different behavior.

\end{appendix}

\end{document}